\documentclass{article}
\usepackage{amsmath, amssymb, amsfonts}
\title{On geodesics in spherical Rindler space} 
\author{Hristu Culetu \footnote{e-mail : hculetu@yahoo.com} \\Ovidius University, Deparment of Physics and Electronics\\ Constanta, Romania\\ORCID: https://orcid.org/0000-0002-3418-2784}

\begin{document}
\numberwithin{equation}{section}
\pagenumbering{arabic}
\maketitle
\newcommand{\fv}{\boldsymbol{f}}
\newcommand{\tv}{\boldsymbol{t}}
\newcommand{\gv}{\boldsymbol{g}}
\newcommand{\OV}{\boldsymbol{O}}
\newcommand{\wv}{\boldsymbol{w}}
\newcommand{\WV}{\boldsymbol{W}}
\newcommand{\NV}{\boldsymbol{N}}
\newcommand{\hv}{\boldsymbol{h}}
\newcommand{\yv}{\boldsymbol{y}}
\newcommand{\RE}{\textrm{Re}}
\newcommand{\IM}{\textrm{Im}}
\newcommand{\rot}{\textrm{rot}}
\newcommand{\dv}{\boldsymbol{d}}
\newcommand{\grad}{\textrm{grad}}
\newcommand{\Tr}{\textrm{Tr}}
\newcommand{\ua}{\uparrow}
\newcommand{\da}{\downarrow}
\newcommand{\ct}{\textrm{const}}
\newcommand{\xv}{\boldsymbol{x}}
\newcommand{\mv}{\boldsymbol{m}}
\newcommand{\rv}{\boldsymbol{r}}
\newcommand{\kv}{\boldsymbol{k}}
\newcommand{\VE}{\boldsymbol{V}}
\newcommand{\sv}{\boldsymbol{s}}
\newcommand{\RV}{\boldsymbol{R}}
\newcommand{\pv}{\boldsymbol{p}}
\newcommand{\PV}{\boldsymbol{P}}
\newcommand{\EV}{\boldsymbol{E}}
\newcommand{\DV}{\boldsymbol{D}}
\newcommand{\BV}{\boldsymbol{B}}
\newcommand{\HV}{\boldsymbol{H}}
\newcommand{\MV}{\boldsymbol{M}}
\newcommand{\be}{\begin{equation}}
\newcommand{\ee}{\end{equation}}
\newcommand{\ba}{\begin{eqnarray}}
\newcommand{\ea}{\end{eqnarray}}
\newcommand{\bq}{\begin{eqnarray*}}
\newcommand{\eq}{\end{eqnarray*}}
\newcommand{\pa}{\partial}
\newcommand{\f}{\frac}
\newcommand{\FV}{\boldsymbol{F}}
\newcommand{\ve}{\boldsymbol{v}}
\newcommand{\AV}{\boldsymbol{A}}
\newcommand{\jv}{\boldsymbol{j}}
\newcommand{\LV}{\boldsymbol{L}}
\newcommand{\SV}{\boldsymbol{S}}
\newcommand{\av}{\boldsymbol{a}}
\newcommand{\qv}{\boldsymbol{q}}
\newcommand{\QV}{\boldsymbol{Q}}
\newcommand{\ev}{\boldsymbol{e}}
\newcommand{\uv}{\boldsymbol{u}}
\newcommand{\KV}{\boldsymbol{K}}
\newcommand{\ro}{\boldsymbol{\rho}}
\newcommand{\si}{\boldsymbol{\sigma}}
\newcommand{\thv}{\boldsymbol{\theta}}
\newcommand{\bv}{\boldsymbol{b}}
\newcommand{\JV}{\boldsymbol{J}}
\newcommand{\nv}{\boldsymbol{n}}
\newcommand{\lv}{\boldsymbol{l}}
\newcommand{\om}{\boldsymbol{\omega}}
\newcommand{\Om}{\boldsymbol{\Omega}}
\newcommand{\Piv}{\boldsymbol{\Pi}}
\newcommand{\UV}{\boldsymbol{U}}
\newcommand{\iv}{\boldsymbol{i}}
\newcommand{\nuv}{\boldsymbol{\nu}}
\newcommand{\muv}{\boldsymbol{\mu}}
\newcommand{\lm}{\boldsymbol{\lambda}}
\newcommand{\Lm}{\boldsymbol{\Lambda}}
\newcommand{\opsi}{\overline{\psi}}
\renewcommand{\tan}{\textrm{tg}}
\renewcommand{\cot}{\textrm{ctg}}
\renewcommand{\sinh}{\textrm{sh}}
\renewcommand{\cosh}{\textrm{ch}}
\renewcommand{\tanh}{\textrm{th}}
\renewcommand{\coth}{\textrm{cth}}

\begin{abstract}
The geodesics in various spherical Rindler frames are investigated. A display of some kinematical quantities of the spacetime is given. The constant acceleration from the metric acts as the surface gravity of the horizon $r = 0$. The radial geodesics are computed both for the Balasubramanian et al. form of the spherical Rindler space and for the non-diagonal metric of Huang and Sun.
 \end{abstract}

\section{Introduction}
As is well-known, Rindler spacetime corresponds to an uniformly accelerating observer in Minkowski space. A less studied case refers to spherical Rindler space, given by a set of radially accelerating observers who are causally disconnected from
a spherical region of radius $b$, located at the origin of Minkowski space \cite{BCCB}, whose boundary represents the horizon of the Rindler space. As Balasubramanian et al. \cite{BCCB} have observed, the thermodynamics of this horizon is subtle, because spherical Rindler space is time-dependent for global reasons: the observers who define it accelerate in different directions. Despite of the time dependence of the metric, Balasubramanian et al. succeeded to compute the gravitational entropy of spherical Rindler space. 

When viewed by a Minkowskian observer, the set of accelerating observers is expanding hyperbolically \cite{HC2}. The authors of \cite{BCCB} concluded that their calculations measure the entanglement entropy between the quantum gravity systems inside and outside of a spherical ball in flat space. On the other hand, the four-dimensional subspace of the Witten bubble geometry \cite{EW} and the spherical Rindler geometry are identical provided the radial coordinate $r$ is large (see also \cite{HC1, HC2}), both of them being Minkowskian. 

  Lee et al. \cite{LP} (see also \cite{UG}) consider the case of true-to-false vacuum phase transition from flat space to de Sitter space using spherical Rindler coordinates inside the bubble. Paithankar and Kolekar \cite{PK} investigated Rindler trajectories in black holes space and Santos \cite{LS} and Moreira \cite{AM} studied thermodynamics properties of fermions in Rindler space (see also \cite{BE, AA}).

 Huang and Sun \cite{HS} also used a spherically symmetric generalization of Rindler spacetime (see also \cite{HC3}) and introduced a new kind of uniformly accelerated reference frame \cite{HG}, with the interesting property that all static observer in uniformly accelerated coordinates have the same constant acceleration, irrespective of their position. In \cite{HS}, the authors have also written down a generalized form of the new uniformly accelerated frame in spherical Rindler coordinates, as in \cite{BCCB}.

Our purpose in this work is to find the equation of geodesics in spherical Rindler spacetime, both for the Balasubramanian et al. and for the Huang and Sun geometries. 

 We work with geometrical units ($c$ = $G$ = 1), unless otherwise specified. 

\section{Spherical Rindler spacetime}
  Spherical Rindler space refers to a family of observers accelerating away from a common center \cite{BCCB}, who are causally disconnected from a spherical region of radius $b$. Starting with Minkowski geometry in spherical coordinates
	   \begin{equation}
		ds^{2} = -d\eta^{2} + d\rho^{2} + \rho^{2}d\Omega^{2},
 \label{2.1}
 \end{equation}
 where $d\Omega^{2}$ stands for the metric on the 2-sphere, $\eta >0$, $\rho >0$, the spherical Rindler coordinates may be defined as \cite{BCCB}
	   \begin{equation}
		r = \sqrt{(\rho - b)^{2} - \eta^{2}},~~~~gt = arctanh \frac{\eta}{\rho - b},
 \label{2.2}
 \end{equation}
 A static observer in spherical Rindler frame at $r = const.$ is moving hyperbolically viewed from Minkowski space. Accelerating observers cannot see the interior of a sphere of radius $b$ having the center at the origin of Minkowski coordinates.
From (2.1) and (2.2) one obtains
	   \begin{equation}
		ds^{2} = -g^{2}r^{2}dt^{2} + dr^{2} +  (b + r cosh~gt)^{2}d\Omega^{2},                                               
 \label{2.3}
 \end{equation}
where $g>0$ is a constant acceleration, being necessary for to get the correct units. The geometry (2.3) is flat, being obtained from the Minkowski metric by a coordinate transformation. However, it does not cover the whole Minkowski space, exactly as the standard Rindler space.

Let us study now the kinematical quantities for a ''static'' observer in the space (2.3), with the velocity vector field
	   \begin{equation}
	u^{a} = \left(\frac{1}{gr}, 0, 0, 0\right), ~~~~u^{a}u_{a} = -1.                                               
 \label{2.4}
 \end{equation}
The corresponding acceleration 4-vector $a^{b} = u^{a}\nabla_{a}u^{b}$ is given by
	   \begin{equation}
	a^{b} = \left(0, \frac{1}{r}, 0, 0\right),~~~~\sqrt{a^{b}a_{b}} = \frac{1}{r}.                                              
 \label{2.5}
 \end{equation}
Since $a^{r}>0$, the gravitational field is attractive.
For the surface gravity corresponding to the horizon $r = 0$ we have
 \begin{equation}
\kappa = \sqrt{a^{b}a_{b}} \sqrt{-g_{tt}}|_{r = 0} = g.
 \label{2.6}
 \end{equation}
In other words, the constant $g$ introduced in (2.3) means the surface gravity $\kappa$.  From the point of view of a Minkowskian observer the horizon corresponds to the light cones $\rho = b \pm \eta$,which are asymptotics for the hyperbolic observer \cite{BCCB}.
The expansion scalar of the ''static'' observer appears as
 \begin{equation}
\Theta \equiv \nabla_{a}u^{a} = \frac{2sinh~gt}{b + rcosh~gt},
 \label{2.7}
 \end{equation}
which is always positive.

The non-zero components of the shear tensor 
 \begin{equation}
\sigma_{ab} = \frac{1}{2}\left(h^{c}_{b}\nabla_{c}u_{a} + h^{c}_{a}\nabla_{c}u_{b}\right) - \frac{1}{3}\Theta h_{ab} + \frac{1}{2}\left(a_{a}u_{b} + a_{b}u_{a}\right)
 \label{2.8}
 \end{equation}
are given by
\begin{equation}
\sigma^{r}_{~r} = -2\sigma^{\theta}_{~\theta} = -2\sigma^{\phi}_{~\phi} = \frac{-2sinh~gt}{3(b + rcosh~gt)},
 \label{2.9}
 \end{equation}
with $\sigma^{a}_{~a} = 0$ and 
\begin{equation}
\sigma^{ab}\sigma_{ab} = \frac{\sqrt{6}}{3} \frac{sinh~gt}{b + rcosh~gt}.
 \label{2.10}
 \end{equation}
 The metric $h_{ab} = g_{ab}+u_{a}u_{b}$ is the projection tensor onto the direction perpendicular to $u_{a}$ and $\sigma_{ab}$ expresses the distorsion of the worldlines in shape without change in volume.

The vorticity tensor $\omega^{a}_{~b}$ is vanishing.

\section{Radial geodesics}
We propose to reach at the geodesic equations with the help of the Lagrangean
	   \begin{equation}
		L = \frac{1}{2}\left(\frac{ds}{d\tau}\right)^{2} = \frac{1}{2}\left(g^{2}r^{2}\dot{t}^{2} - \dot{r}^{2} - (b + r cosh~gt)^{2}(\dot{\theta}^{2} + sin^{2}\theta ~\dot{\phi}^{2})\right),                                              
 \label{3.1}
 \end{equation}
where $\tau$ is the proper time. One looks, for simplicity, for the radial geodesics, namely, $\dot{\theta} = \dot{\phi} = 0$. Therefore, from (3.1) we get
	   \begin{equation}
		 g^{2}r^{2}\dot{t}^{2} - \dot{r}^{2} = 1.                                              
 \label{3.2}
 \end{equation}
The Euler-Lagrange equations read
	   \begin{equation}
	\frac{\partial L}{\partial x^{a}} - \frac{d}{d\tau}\frac{\partial L}{\partial \dot{x}^{a}} = 0.	
 \label{3.3}
 \end{equation}
 For $\theta, \phi = const.$ the geometry (2.3) is static and, from (3.3) with $a =t$ we get 
 \begin{equation}
\dot{t} = \frac{E}{g^{2}r^{2}}, 
 \label{3.4}
 \end{equation}
where $E$ is the energy per unit mass of the particle and $\dot{t} = dt/d\tau$. The case $a = r$ yields
  \begin{equation}
\ddot{r} = -g^{2}r\dot{t}^{2}. 
 \label{3.5}
 \end{equation}
From (3.2), (3.4) and (3.5) one easily obtains
  \begin{equation}
\dot{r} = - \frac{g\tau}{\sqrt{E^{2} - g^{2}\tau^{2}}},~~~ \dot{t} = \frac{E}{E^{2} - g^{2}\tau^{2}},~~~\tau < \frac{E}{g}.  
 \label{3.6}
 \end{equation}
Getting rid of $\tau$ one obtains $\dot{r}$ and $\dot{t}$ in terms of $r$
  \begin{equation}
\dot{r} = - \frac{\sqrt{E^{2} - g^{2}r^{2}}}{gr},~~~~ \dot{t} = \frac{E}{g^{2}r^{2}}, 
 \label{3.7}
 \end{equation}
It is an easy task to check that the 4-velocity
  \begin{equation}
u^{b} = \left(\frac{E}{g^{2}r^{2}}, -\frac{\sqrt{E^{2} - g^{2}r^{2}}}{gr}, 0, 0\right)
 \label{3.8}
 \end{equation}
yields $a^{b} = u^{a}\nabla_{a}u^{b} = 0$; namely, $u^{a}$ is tangent at the timelike geodesic.
From (3.6) we have
  \begin{equation}
r(\tau) = \frac{1}{g}\sqrt{E^{2} - g^{2}\tau^{2}},~~~ t(\tau) = \frac{1}{2g}~ln\frac{E + g\tau}{E - g\tau}.  
 \label{3.9}
 \end{equation}
We get rid of $\tau$ from (3.9) and obtain the radial equation of motion
\begin{equation}
r(t) = \frac{E}{g~cosh~gt}, 
 \label{3.10}
 \end{equation}
with appropriate initial conditions: $r(0) = E/g$ and $r \rightarrow 0$ when $t \rightarrow \infty$, i.e. the horizon is reached after an infinite time. For the velocity of the particle one obtains
\begin{equation}
\frac{dr}{dt} = -\frac{E~tanh~gt}{cosh~gt}, 
 \label{3.11}
 \end{equation}
with $|dr/dt|<1$, as it should be for a massive particle.

As far as the null radial geodesics are concerned, they are obtained from $L = 0$
	   \begin{equation}
		 g^{2}r^{2}\dot{t}^{2} - \dot{r}^{2} = 0,                                              
 \label{3.12}
 \end{equation}
giving us $r(t) = (1/g)e^{-gt}$, with $r(0) = 1/g$ and $r \rightarrow 0$ at infinity.  The velocity $v = dr/dt = -e^{-gt}$ shows that the null particle starts with unit velocity and has $v = 0$ when $t \rightarrow \infty$.

\section{Geodesics in Huang-Sun spacetime}
Huang and Sun \cite{HS} investigated a new form of the flat geometry related to the Rindler metric, in spherical coordinates. Starting with the standard Minkowski metric (2.1) and performing the coordinate transformation
	   \begin{equation}
		\eta = \frac{1}{g}~sinh~gt,~~~~\rho = r + \frac{1}{g}~(cosh~gt - 1)
 \label{4.1}
 \end{equation}
they obtained
	\begin{equation}
			ds^{2} = -dt^{2} + 2sinh~gt~dtdr + dr^{2} + \left[r + \frac{1}{g}~(cosh~gt - 1)\right]^{2}d\Omega^{2}.                                      \label{4.2}
 \end{equation}
Huang and Sun already computed the radial null geodesics for (4.2). We look now for the timelike radial geodesics, obtained from  
	\begin{equation}
			ds^{2} = -dt^{2} + 2sinh~gt~dtdr + dr^{2}                                                                                         \label{4.3}
 \end{equation}
One equation is given by
	   \begin{equation}
		 \dot{t}^{2} - 2sinh~gt~\dot{t}\dot{r} - \dot{r}^{2} = 1,                                              
 \label{4.4}
 \end{equation}
and the others are obtained from the Euler-Lagrange equations
 \begin{equation}
 \dot{r} + \dot{t}sinh~gt = 0,
 \label{4.5}
 \end{equation} 
and
\begin{equation}
\ddot{t} - \ddot{r}sinh~gt = 0
 \label{4.6}
 \end{equation}
Noting that in (4.5) the constant of integration w.r.t. $\tau$ has been chosen zero, to be in accordance with (4.4). Eq.(4.5) gives us
	   \begin{equation}
		r(t) = \frac{1}{g}(\alpha - cosh~gt),~~~~t<\frac{1}{g}ln~(\alpha + \sqrt{\alpha^{2} - 1}),
 \label{4.7}
 \end{equation}
where $\alpha$ is a constant of integration. From (4.5) and (4.7) we get
	   \begin{equation}
		\dot{r} = -tanh~gt,~~~~\dot{t} = \frac{1}{cosh~gt}.
 \label{4.8}
 \end{equation}
By means of (4.8), it is clear that (4.6) is satisfied.
Using (4.8) one finds that $a^{b} = u^{a}\nabla_{a}u^{b} = 0$, where $u^{a} = (1/cosh~gt, -tanh~gt, 0, 0)$. In other words, $r(t)$ from (4.7) indeed represents the geodesic trajectory.

\section{Conclusions}
The spherical Rindler spacetime is less studied compared to the standard Rindler geometry, written in Rindler or Moller forms. The boundary of the Rindler space is given by its horizon at $r = 0$. The radially accelerated observer are disconnected from a spherical region of radius $b$, centered at the origin of Minkowski space. 

The 4-dimensional subspace of the Witten bubble geometry and the spherical Rindler geometry are identical provided the radial coordinate is large. The horizon at $r = 0$ is the boundary of the sphere of radius $b$. The main goal of the paper was to calculate radial geodesics for the two versions of the Rindler space mentioned above. It is interesting that all geodesics are expressed in a simple and exact form, even though the metric (2.3) is time dependent. An advantage of the spherically symmetric Rindler coordinates compared to the usual Cartesian form is related to its finite horizon, as for the de Sitter space, and that may be connected to Cosmology, in spite that (2.3) is a flat metric. The system of spherically accelerating observers moving hyperbolically and having a horizon behind them can be applied to fast expanding fluids and eventually to acoustic black holes. \\

\end{document}